\pdfoutput=1  % must appear within the first 5 lines: forces pdfLaTeX (figures are PNG)
\documentclass[11pt]{article}

\usepackage[margin=1in]{geometry}
\usepackage{amsmath,amssymb,amsfonts}
\usepackage{graphicx}
\graphicspath{{figures/}}
\usepackage{booktabs}
\usepackage{algorithm}
\usepackage{algpseudocode}
\usepackage{caption}
\usepackage{url}
\usepackage[hidelinks]{hyperref}
\usepackage{placeins}

\let\oldsection\section
\renewcommand{\section}{\FloatBarrier\oldsection}

\newcommand{\ket}[1]{\lvert #1 \rangle}
\newcommand{\bra}[1]{\langle #1 \rvert}
\newcommand{\expv}[1]{\langle #1 \rangle}

\title{\bfseries Variance-Reduced Trajectory Unravelings for GPU Noisy Quantum-Circuit
Simulation: Characterization and a Qiskit-Aer Integration Gap}

\author{Chun-Yeol You\thanks{Department of Physics and Chemistry, DGIST, Daegu 42988,
Republic of Korea. ORCID: 0000-0001-9549-8611. E-mail: \texttt{cyyou@dgist.ac.kr}}}

\date{\today}

\begin{document}
\maketitle

\begin{abstract}
Monte-Carlo trajectory (quantum-jump) methods are the practical route to simulating noisy
quantum circuits once the exact density-matrix method is precluded by its $4^n$ memory cost.
Their bottleneck is estimator variance: resolving one expectation value can demand thousands of
trajectories. Recent tensor-network work shows that \emph{variance-reduced unravelings}---projector
and analog sampling---sharply cut this variance, but only on CPU matrix-product-state backends,
with no path into production tooling. We implement both unravelings on a \emph{GPU
dense-statevector} trajectory engine and validate them against the exact density matrix
(ideal-circuit fidelity $1-2.2\times10^{-16}$; $1/\sqrt{N}$ convergence; all unravelings unbiased
to trace distance $<0.01$). On a single consumer GPU, projector unraveling reaches a target
standard error with $20.8\times$ fewer trajectories than Qiskit-Aer's
\texttt{batched\_shots\_gpu} at $n=10$, a factor that holds at
$19$--$26\times$ across $n=8$--$20$. A regime map places analog sampling optimal at weak noise and
projector at strong noise, crossing near $\gamma t\approx0.35$. We further report a systems
finding: Qiskit-Aer applies noise at the \emph{channel} level and reconstructs a canonical Kraus
decomposition at apply time, discarding any user-supplied unraveling, so variance-reduced
unravelings cannot be delivered through its public API. Because Aer's Born-rule collapse machinery
already exists, we specify a minimal change that would unlock the technique in production.
\end{abstract}


\vspace{1em}

\section{Introduction}
Classical simulation of quantum circuits underpins algorithm development,
compiler validation, and the assessment of error-mitigation strategies in the
noisy intermediate-scale quantum (NISQ) era \cite{preskill2018}. For \emph{ideal} circuits,
statevector simulators store $2^n$ complex amplitudes and have been pushed to tens of qubits
on GPUs and supercomputers \cite{haner2017,quest2019,intelqs2020,cuquantum2023,qiskit_aer}.
The far more demanding---and more practically relevant---task is to simulate circuits
\emph{with realistic noise}, since the noise is precisely what separates a NISQ device from
an ideal one and what error-mitigation methods must model
\cite{temme2017,sparse_pauli_lindblad2023}.

The textbook route to noise is to evolve the density matrix $\rho$, but its $4^n$ memory
footprint caps it at roughly $n\!\approx\!14$ qubits on an $8$\,GB GPU and $n\!\approx\!16$
even on an $80$\,GB datacenter GPU---far short of the ideal-statevector reach. The scalable
alternative is the Monte-Carlo \emph{trajectory} (quantum-jump / Monte-Carlo wave-function,
MCWF) method \cite{dalibard1992,molmer1993,plenio1998,breuer_petruccione,daley2014}: the
open-system dynamics is ``unraveled'' into an ensemble of pure statevector trajectories, and
observables are estimated by averaging over them. Trajectory methods inherit the ideal
statevector's $2^n$ memory, so they reach far larger $n$ than the density matrix; their cost
is instead \emph{statistical}, set by the estimator variance. Resolving an expectation value
to a target standard error can require thousands of trajectories, and this multiplies the
already-large cost of the many circuit evaluations demanded by variational algorithms and
error mitigation.

A key but underexploited fact is that a given open-system evolution admits \emph{many}
unravelings that reproduce the same $\rho$ while differing greatly in estimator variance
\cite{breuer_petruccione,daley2014}. Recent tensor-network work \cite{tjm2026} makes this
concrete, introducing \emph{projector} and \emph{analog} unravelings with provably lower
variance and demonstrating them for sparse Pauli-Lindblad noise
\cite{sparse_pauli_lindblad2023}. That work, however, is realized on CPU
matrix-product-state (MPS) backends \cite{vidal2003,cirac2021mps} and is not available in any
production simulator; related tensor-network noisy-simulation methods
\cite{lpdo2023,mpdo_prr,mpdo_tomography2025,noh2020} share both properties. Conversely,
production GPU simulators \cite{qiskit_aer,cuquantum2023} accelerate only the \emph{standard}
trajectory unraveling and expose no variance-reduced alternative. Variance-reduced
unravelings on a GPU dense statevector---the representation most simulators and hardware
vendors optimize---are thus unexplored, and their path into production tooling is unknown.

\textbf{Contributions.} This paper occupies exactly that gap through four connected contributions.
First, we implement projector and analog unravelings on a \emph{GPU dense-statevector} trajectory engine
and validate them against Qiskit-Aer's exact density-matrix method, obtaining ideal-circuit fidelity
$1-\mathcal{O}(10^{-16})$ and $1/\sqrt{N}$ convergence of the reconstructed $\rho$
(Section~\ref{sec:method},~\ref{sec:validation}). Building on that validated engine, we then characterize
the variance reduction quantitatively: we measure a $\sim\!21\times$ reduction in the trajectories needed
to reach a target standard error in the strong-noise regime, show that this factor is stable across
$n=8$--$20$, and construct a noise-strength regime map whose analog/projector crossover falls near
$\gamma t\approx0.35$ (Section~\ref{sec:results}); the same analysis reveals how the advantage depends on the
choice of observable (GHZ stabilizers) and how interleaved coherent dynamics erodes it. To place these
numbers on a production footing, we next benchmark head-to-head against Qiskit-Aer's
\texttt{batched\_shots\_gpu} path, verifying that our standard unraveling is statistically identical to
Aer's---an apples-to-apples control that attributes the gain to the unraveling rather than to our
implementation---while the projector unraveling attains the same accuracy with $\sim\!21\times$ fewer
trajectories (Section~\ref{sec:aer}). Finally, and unexpectedly, we report and diagnose an \emph{integration
gap}: because Qiskit-Aer applies noise at the channel level and reconstructs a canonical Kraus
decomposition at apply time, it discards any user-supplied unraveling, so variance-reduced unravelings
cannot be delivered through its public API. Since the Born-rule collapse machinery nevertheless already
exists inside Aer, we are able to specify the minimal change that would unlock the technique in
production (Section~\ref{sec:aer}).
Our framing deliberately avoids a raw-throughput competition with cuStateVec-class kernels
\cite{cuquantum2023} or GPU circuit-partitioning systems \cite{atlas_sc24,questab_ics25};
the contribution is a statistical technique, its rigorous GPU characterization, and its
precise production-integration path. All results use a single consumer GPU with fixed seeds
and accompanying raw data, and we project the datacenter-GPU implications in
Section~\ref{sec:limitations}.

\section{Qiskit, Its Physical Significance, and Recent Trends}\label{sec:qiskit}

\begin{figure}[!htbp]\centering
\includegraphics[width=\linewidth]{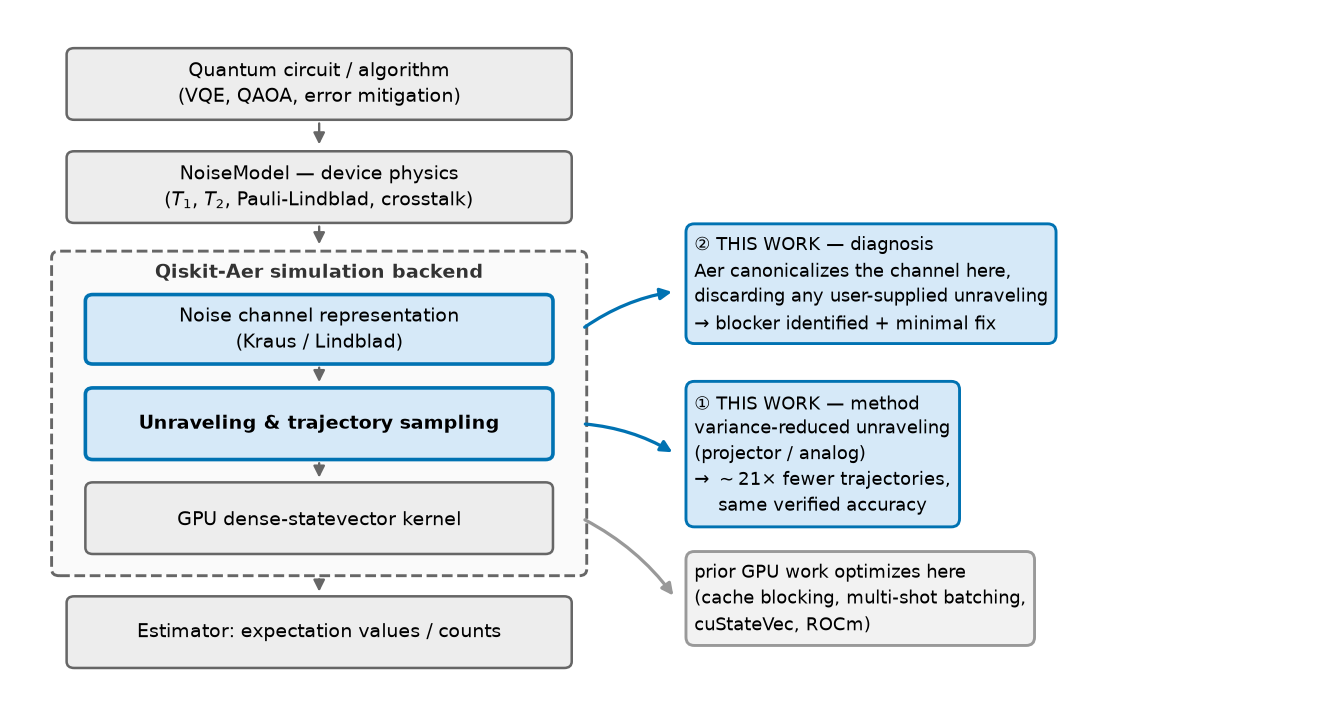}
\caption{Where this work intervenes in the Qiskit-Aer stack. Left: the simulation pipeline. A quantum
circuit and a \texttt{NoiseModel} carrying measured device physics ($T_1$, $T_2$, Pauli-Lindblad rates,
crosstalk) are passed to the Qiskit-Aer backend (dashed box), which (i) represents the noise as a
channel (Kraus/Lindblad), (ii) \emph{unravels} that channel into sampled quantum-jump trajectories, and
(iii) evolves each trajectory on a GPU dense-statevector kernel, finally producing expectation values or
counts. Right: the two contributions of this paper, both localized to the highlighted blue stages.
\textcircled{\scriptsize 1} \emph{Method}---we replace the standard Pauli-flip unraveling at the
sampling stage with variance-reduced projector/analog unravelings, obtaining $\sim\!21\times$ fewer
trajectories at identical, independently verified accuracy. \textcircled{\scriptsize 2}
\emph{Diagnosis}---we show that Aer canonicalizes the noise channel at the representation stage,
discarding any user-supplied unraveling, which is precisely why \textcircled{\scriptsize 1} cannot be
delivered through the public API today; we identify the minimal change that removes this blocker. For
contrast, essentially all prior GPU work on Aer (cache blocking, multi-shot batching, cuStateVec, ROCm)
optimizes the kernel stage (grey arrow), i.e.\ the cost \emph{per} trajectory---an axis orthogonal to,
and multiplicative with, the reduction in the \emph{number} of trajectories reported here.}
\label{fig:concept}
\end{figure}

\subsection{Why Qiskit matters physically}
Qiskit \cite{qiskit_aer,qiskit2024,wille2019qiskit} is not merely a programming convenience: it is the
de-facto bridge between the \emph{physics of a real device} and the algorithms run on it. Its
\texttt{NoiseModel} layer encodes measured device physics---$T_1$ amplitude damping, $T_2$ dephasing,
gate-dependent depolarizing rates, readout error, and, increasingly, correlated crosstalk expressed as
sparse Pauli-Lindblad generators learned directly from hardware \cite{sparse_pauli_lindblad2023,
georgopoulos2021modeling}. Because these generators are the same Lindbladians that govern the
laboratory system \eqref{eq:lindblad}, a Qiskit noisy simulation is, in effect, a numerical open-quantum-system
experiment. This gives Qiskit a specific physical role that a bare linear-algebra simulator does not have.
It serves, first, as a vehicle for \emph{predictive device modeling}, since comparing simulated with
measured observables is how one validates a noise model and thus one's physical understanding of the
device \cite{georgopoulos2021modeling,nation2023suppressing}. It serves, second, as a \emph{control
experiment for quantum-advantage claims}: classical simulation is the yardstick against which hardware
results are judged, and several ``advantage'' claims have been narrowed or overturned by improved
classical simulation \cite{arute2019supremacy,pan2022sycamore,markov2008tensor,gray2021hyper}. It serves,
third, as the \emph{engine of error mitigation}, because zero-noise extrapolation and probabilistic error
cancellation \cite{temme2017,kandala2019mitigation,endo2021mitigation} require accurate noisy expectation
values, and the recent ``utility'' experiments \cite{kim2023utility} were assessed against precisely such
simulations.
Consequently, the cost of noisy simulation in Qiskit is not an implementation detail---it directly limits
what device physics can be modeled, and at what scale.

\subsection{Recent research trends}
Four trends frame the present work. \emph{(i) Noise-aware NISQ algorithms.} Variational algorithms and
their error-mitigated variants \cite{cerezo2021vqa,bharti2022nisq,endo2021mitigation} consume enormous
numbers of \emph{noisy expectation-value estimates}; the per-estimate statistical cost therefore dominates
the total budget. \emph{(ii) Classical simulation as a moving target.} Tensor-network contraction
\cite{markov2008tensor,gray2021hyper,pan2022sycamore}, stabilizer-rank
\cite{aaronson2004stabilizer,bravyi2016clifford}, and MPS methods \cite{vidal2003,schollwock2011dmrg,
cirac2021mps,zhou2020} keep raising the bar for what is classically reachable. \emph{(iii) Noise as a
resource for classical tractability.} Noise limits entanglement growth, making noisy circuits
\emph{easier} to simulate approximately \cite{noh2020,zhou2020}---the insight underlying matrix-product
density operators \cite{verstraete2004mpdo,zwolak2004mixed,mpdo_prr,lpdo2023,mpdo_tomography2025} and,
most recently, variance-reduced trajectory unravelings \cite{tjm2026}. \emph{(iv) Hardware acceleration.}
GPU statevector engines \cite{cuquantum2023,quest2019,intelqs2020,haner2017}, cache blocking
\cite{cacheblocking2020}, multi-shot batching \cite{efficient_multishot2023}, ROCm portability
\cite{rocm_qiskit2024}, and multi-GPU partitioning \cite{atlas_sc24,questab_ics25,multigpu_network2025}
have all been pursued. Trends (i)+(iii) create the demand our method serves; trend (iv) supplies the
platform; yet (iii) and (iv) have so far advanced in separate communities.

\subsection{Prior Qiskit-related simulation research and what we improve}
Work targeting Qiskit-Aer has, to date, optimized the \emph{cost per amplitude or per shot}:
cache blocking to fit large statevectors across memory hierarchies \cite{cacheblocking2020}, GPU
multi-shot batching for noisy circuits \cite{efficient_multishot2023}, cuStateVec kernel offload
\cite{cuquantum2023}, ROCm/AMD portability \cite{rocm_qiskit2024}, and device-variability-aware
transpilation \cite{nation2023suppressing}. All of these accelerate or rearrange the \emph{same}
standard (Pauli-flip) unraveling of the noise.

We improve a different and previously untouched axis: the \emph{statistical efficiency of the unraveling
itself}. Our contribution is complementary---it multiplies with, rather than competes against, every
optimization above---and consists of two concrete improvements over the current Qiskit-Aer state of the
art. The first is \emph{fewer trajectories for the same accuracy}: replacing the standard unraveling by
the projector unraveling reduces the trajectories needed to reach a target standard error by
$\sim\!21\times$ in the strong-noise regime (Section~\ref{sec:aer}), at identical, verified accuracy. The
second is \emph{removing an architectural blocker}: we show \emph{why} Qiskit-Aer cannot express such
unravelings today---it canonicalizes the noise channel at apply time---and specify the minimal change
that would enable them, reusing collapse machinery Aer already possesses.

\subsection{Differentiation of this work}
Table~\ref{tab:diff} summarizes how this work differs from the closest lines of research. In one sentence:
prior variance-reduction work is CPU/tensor-network and unintegrated; prior GPU work is
production-integrated but statistically standard; \emph{this work is the first to place a
variance-reduced unraveling on a GPU dense statevector, quantify it against a production baseline, and
identify the concrete production-integration path.}

\begin{table}[!htbp]
\caption{Differentiation from the closest related work.}
\label{tab:diff}
\centering

\begin{tabular}{@{}lccc@{}}
\toprule
 & variance- & GPU dense & Aer \\
 & reduced & statevector & analysis \\
\midrule
Qiskit-Aer \cite{qiskit_aer,efficient_multishot2023} & no & yes & baseline \\
cuStateVec \cite{cuquantum2023}                      & no & yes & no \\
Atlas/quEStab \cite{atlas_sc24,questab_ics25}        & no & ideal & no \\
MPDO/LPDO \cite{mpdo_prr,lpdo2023}                   & n/a & no & no \\
Tensor jump \cite{tjm2026}                           & yes & no & no \\
\textbf{This work}                                   & \textbf{yes} & \textbf{yes} & \textbf{yes} \\
\bottomrule
\end{tabular}
\end{table}

\section{Background}
\subsection{Open-system dynamics and the MCWF unraveling}
A Markovian open-system evolution is generated by a Lindblad master equation
\cite{lindblad1976,gks1976,breuer_petruccione,nielsen_chuang}
\begin{equation}
\dot\rho \;=\; \mathcal{L}(\rho) \;=\; \sum_m \gamma_m\!\left(
L_m\,\rho\,L_m^\dagger - \tfrac{1}{2}\{L_m^\dagger L_m,\rho\}\right),
\label{eq:lindblad}
\end{equation}
with jump operators $L_m$ and rates $\gamma_m\ge 0$. For the sparse Pauli-Lindblad model
\cite{sparse_pauli_lindblad2023} the $L_m=P_m$ are Pauli strings ($P_m^\dagger P_m=\mathbb{1}$).

The Monte-Carlo wave-function (MCWF) method \cite{dalibard1992,molmer1993,plenio1998,daley2014}
unravels \eqref{eq:lindblad} into pure-state trajectories. Over an interval $\delta t$ with no
coherent Hamiltonian, a trajectory $\ket{\psi}$ evolves under the non-Hermitian effective
Hamiltonian $H_{\mathrm{eff}}=-\tfrac{i}{2}\sum_m \gamma_m L_m^\dagger L_m$, giving an
unnormalized state $\ket{\tilde\psi}=e^{-iH_{\mathrm{eff}}\delta t}\ket{\psi}$ with jump
probability
\begin{equation}
\delta p \;=\; 1-\langle\tilde\psi|\tilde\psi\rangle
\;=\; 1-\Big\langle\psi\Big|\,e^{-\delta t\sum_m\gamma_m L_m^\dagger L_m}\Big|\psi\Big\rangle .
\label{eq:jumpprob}
\end{equation}
With probability $1-\delta p$ no jump occurs and $\ket{\psi}\!\leftarrow\!\ket{\tilde\psi}/\|\tilde\psi\|$;
otherwise a channel $m$ is drawn with probability $p_m\propto\gamma_m\|L_m\ket{\psi}\|^2$ and
$\ket{\psi}\!\leftarrow\! L_m\ket{\psi}/\|L_m\ket{\psi}\|$. Averaging
$\ket{\psi}\!\bra{\psi}$ over trajectories reproduces $\rho$ exactly. For Pauli jumps
$L_m^\dagger L_m=\mathbb{1}$, so \eqref{eq:jumpprob} reduces to a state-independent rate
$\Gamma=\sum_m\gamma_m$ and $\delta p = 1-e^{-\Gamma\delta t}$, which we sample \emph{exactly}
(as a Poisson process) rather than time-stepping.

\subsection{Unraveling freedom and variance}
A given generator $\mathcal{L}$ is invariant under
$L_m\mapsto \sum_k u_{mk} L_k$ for any isometry $u$ \cite{breuer_petruccione}. Different choices
therefore reproduce the same $\rho$ (unbiased) but yield estimators of an observable $O$ with
different variance $\mathrm{Var}[\hat O]=\tfrac{1}{N}\big(\mathbb{E}[\expv{O}^2]-\mathbb{E}[\expv{O}]^2\big)$.
We use three unravelings of a single dephasing channel $\gamma(Z\rho Z-\rho)$, with
$\Pi_\pm=(\mathbb{1}\pm Z)/2$:
\begin{align}
\text{standard:}\quad & L = \sqrt{\gamma}\,Z, \label{eq:std}\\
\text{projector:}\quad & L_\pm = \sqrt{2\gamma}\,\Pi_\pm, \label{eq:proj}\\
\text{analog:}\quad & L_\theta = \sqrt{\lambda\,w(\theta)}\,e^{i\theta Z},\;\;
\lambda\,\mathbb{E}_w[\sin^2\theta]=\gamma. \label{eq:analog}
\end{align}
\emph{Projector} unraveling reproduces the same generator: with $L_\pm$ from \eqref{eq:proj},
$\sum_\pm L_\pm\rho L_\pm^\dagger = \gamma(\rho + Z\rho Z)$ and $\sum_\pm L_\pm^\dagger L_\pm=2\gamma\mathbb{1}$,
so the dissipator equals $\gamma(Z\rho Z-\rho)$, identical to \eqref{eq:std}. A jump now
\emph{collapses} the state onto a $Z$-eigenspace. \emph{Analog} unraveling applies frequent
near-identity kicks $e^{i\theta Z}=\cos\theta\,\mathbb{1}+i\sin\theta\,Z$, matched to $\gamma$
through $\lambda\,\mathbb{E}_w[\sin^2\theta]=\gamma$.

\subsection{Why the variance differs}
Consider an observable $O$ that anticommutes with the active channels, $\{O,Z\}=0$, prepared in
an eigenstate of $O$ (e.g.\ $O=X$, $\ket{+}$), under pure dephasing with no coherent evolution.
The exact mean decays as $\expv{O}_t=e^{-2\gamma t}$. Under \emph{standard} unraveling each jump
applies $Z$, flipping $\ket{+}\!\leftrightarrow\!\ket{-}$, so $\expv{O}=(-1)^{N(t)}$ with
$N(t)\sim\mathrm{Poisson}(\gamma t)$; the estimator is a $\pm1$ telegraph with
\begin{equation}
\mathrm{Var}_{\mathrm{std}} = 1-e^{-4\gamma t}.
\label{eq:varstd}
\end{equation}
Under \emph{projector} unraveling the first jump collapses $\ket{+}$ onto a $Z$-eigenstate, where
$\{O,Z\}=0$ forces $\expv{O}=0$ permanently---an ``absorbing window'' \cite{tjm2026}. The
estimator is then Bernoulli ($1$ if no jump in $[0,t]$, else $0$), giving
\begin{equation}
\mathrm{Var}_{\mathrm{proj}} = e^{-2\gamma t}\big(1-e^{-2\gamma t}\big),
\label{eq:varproj}
\end{equation}
so $\mathrm{Var}_{\mathrm{proj}}/\mathrm{Var}_{\mathrm{std}} = e^{-2\gamma t}/(1+e^{-2\gamma t}) \to 0$
in the strong-noise limit. Analog unraveling instead suppresses per-trajectory fluctuations by
replacing rare large flips with frequent small kicks, and is most effective at weak noise. These
predictions are the theoretical anchors we reproduce empirically in Section~\ref{sec:results}.

\section{Method: GPU dense-statevector engine}\label{sec:method}
\subsection{State and gate application}
We store an $n$-qubit statevector as a rank-$n$ complex tensor $\psi\in\mathbb{C}^{2\times\cdots\times2}$
on the GPU (cupy), with axis $q$ indexing qubit $q$. A single-qubit gate $U\in\mathbb{C}^{2\times2}$
on qubit $q$ is a tensor contraction along axis $q$,
\begin{equation}
\psi' = \mathrm{moveaxis}\big(U \cdot_{q}\, \psi,\, 0\!\to\! q\big),\qquad
\psi'_{\ldots i_q \ldots} = \sum_{j} U_{i_q j}\,\psi_{\ldots j \ldots},
\end{equation}
and a two-qubit gate $U\in\mathbb{C}^{4\times4}$ (reshaped to $2^{\times4}$) contracts along its
two qubit axes. We read out a little-endian flat statevector for comparison with Qiskit
conventions. Each gate touches all $2^n$ amplitudes once, so gate application is
memory-bandwidth-bound---the regime in which GPUs excel and in which the per-trajectory cost
scales as $\mathcal{O}(2^n)$.

\subsection{Unraveling operations}
For a dephasing channel on qubit $q$ we implement \eqref{eq:std}--\eqref{eq:analog} as:
\emph{standard}, apply $Z_q$ on a jump; \emph{projector}, measure
$\langle Z_q\rangle$, form $p_\pm=\tfrac{1}{2}(1\pm\langle Z_q\rangle)$, draw a sign, and
collapse
\begin{equation}
\psi \leftarrow \tfrac{1}{2}\big(\psi \pm Z_q\!\cdot\!\psi\big)\big/\big\|\tfrac{1}{2}(\psi \pm Z_q\!\cdot\!\psi)\big\|;
\end{equation}
\emph{analog}, apply $e^{i\theta Z_q}=\cos\theta\,\mathbb{1}+i\sin\theta\,Z_q$ with a two-point
sign law $\theta\in\{\pm\theta_0\}$ so that $\lambda\sin^2\theta_0=\gamma$. The expectation
$\langle O_q\rangle=\mathrm{Re}\,\psi^\dagger (O_q\!\cdot\!\psi)$ and the norm are GPU reductions;
the collapse is an elementwise combination---all standard, bandwidth-bound primitives. For a
global observable such as the GHZ stabilizer $X^{\otimes n}$ we use
$\langle X^{\otimes n}\rangle = \mathrm{Re}\sum_x \psi_x^{*}\,\psi_{\bar x}$, where
$\bar x = x \oplus (2^n\!-\!1)$ is the bitwise complement.

\subsection{Exact event sampling and the estimator}
Because Pauli-Lindblad rates are state-independent (Sec.~II-A), we sample noise \emph{events}
analytically rather than time-stepping, eliminating discretization error. Over an idle window of
duration $\tau$ on channel $q$: standard draws $k\sim\mathrm{Poisson}(\gamma\tau)$ and applies
$Z_q^{\,k}$ (parity); projector jumps with probability $1-e^{-2\gamma\tau}$ and collapses once
(subsequent jumps within the window leave a $Z$-eigenstate invariant); analog draws
$k\sim\mathrm{Poisson}(\lambda\tau)$ signed kicks. Layered circuits interleave gate stages with
such windows. Each trajectory returns a per-trajectory expectation $\expv{\psi|O|\psi}$, and the
estimator is the sample mean over $N$ trajectories,
$\hat O = \tfrac{1}{N}\sum_{i=1}^{N}\expv{\psi^{(i)}|O|\psi^{(i)}}$, with standard error
$\mathrm{SE}=\sqrt{\mathrm{Var}[\expv{O}]/N}$. Algorithm~\ref{alg:traj} summarizes one trajectory.

\begin{algorithm}[t]
\caption{One trajectory (layered circuit, unraveling $\in\{$std, proj, analog$\}$)}
\label{alg:traj}
\begin{algorithmic}[1]
\State $\psi \leftarrow \ket{0}^{\otimes n}$
\For{each stage in circuit}
  \If{gate stage} apply each gate to $\psi$ by contraction
  \Else{ \textbf{(noise window $\tau$)}}
    \For{each qubit $q$ with an active channel}
      \State sample events per the unraveling and update $\psi$
    \EndFor
  \EndIf
\EndFor
\State \Return $\expv{\psi|O|\psi}$
\end{algorithmic}
\end{algorithm}

\section{Validation}\label{sec:validation}
\begin{figure}[!htbp]\centering
\includegraphics[width=\linewidth]{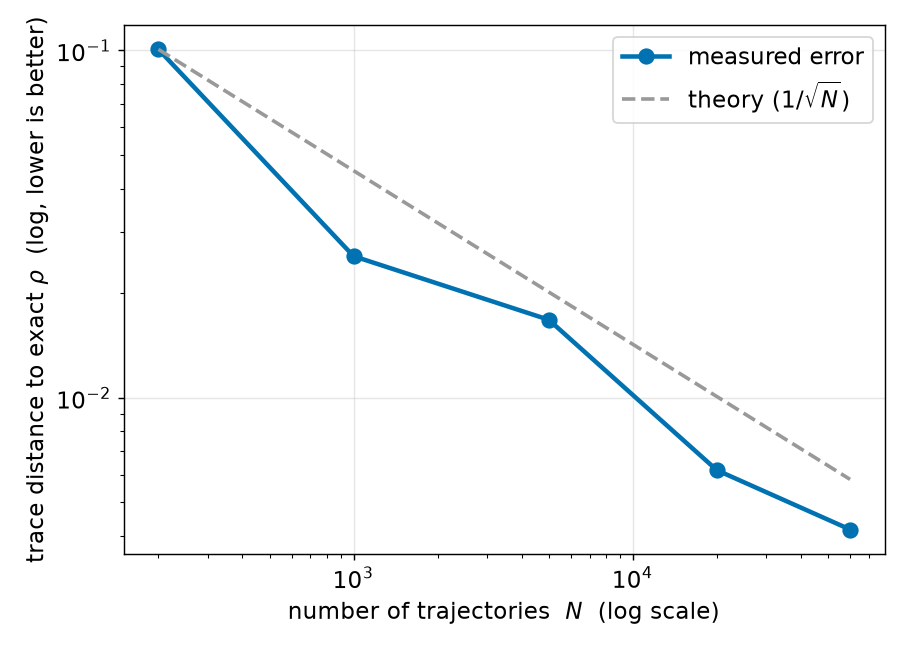}
\caption{Correctness of the GPU trajectory engine. We reconstruct the density matrix
$\rho=\frac1N\sum_i\ket{\psi^{(i)}}\!\bra{\psi^{(i)}}$ from $N$ standard-unraveling trajectories and
plot its trace distance to Qiskit-Aer's \emph{exact} \texttt{density\_matrix} result, on log--log
axes; the $x$-axis is the number of trajectories $N$ ($2\times10^2$ to $6\times10^4$) and the
$y$-axis is the trace distance (lower is more accurate). Blue circles are the measured error; the
grey dashed line is the theoretical Monte-Carlo rate $\propto 1/\sqrt{N}$ anchored at the first
point. The measured error falls from $0.101$ to $4.2\times10^{-3}$ and lies on the $1/\sqrt{N}$
reference throughout (the product $D\!\cdot\!\sqrt{N}$ stays near $1.0$), so the estimator is
unbiased and converges at the statistically optimal rate---establishing that ``fewer trajectories''
(Figs.~\ref{fig:regime}--\ref{fig:aer}) is a genuine efficiency gain, not a loss of accuracy. System:
$6$-qubit circuit, depolarizing noise on single-qubit gates, reference purity $\mathrm{Tr}\,\rho^2=0.61$.}
\label{fig:conv}
\end{figure}

We validate correctness at three levels against Qiskit-Aer references.
\emph{(i) Ideal circuits.} For an $8$-qubit entangling circuit the engine reproduces
\texttt{qiskit.quantum\_info.Statevector} to fidelity $1-2.2\times10^{-16}$, and the GPU path of
Aer likewise matches the reference to $1-4.4\times10^{-16}$---machine precision, confirming the
gate contractions and little-endian readout.
\emph{(ii) Convergence to the exact density matrix.} For a $6$-qubit circuit with depolarizing
noise on single-qubit gates, we reconstruct $\rho=\tfrac{1}{N}\sum_i\ket{\psi^{(i)}}\!\bra{\psi^{(i)}}$
and compare to Aer's exact \texttt{density\_matrix} (purity $\mathrm{Tr}\,\rho^2=0.6143$). The
trace distance falls from $0.1007$ at $N=200$ to $4.17\times10^{-3}$ at $N=6\times10^4$; the
product $D\!\cdot\!\sqrt{N}$ stays near constant ($1.42,0.81,1.19,0.87,1.02$ across
$N=2\times10^2$--$6\times10^4$), matching the $1/\sqrt{N}$ Monte-Carlo rate (Fig.~\ref{fig:conv}).
\emph{(iii) Unbiasedness of every unraveling.} For an $8$-qubit GHZ preparation followed by
dephasing, all three unravelings reproduce Aer's exact $\rho$ at $N=1.5\times10^4$: trace distance
$0.0093$ (standard), $0.0045$ (projector), $0.0063$ (analog)---each $<0.01$---so the variance
reduction changes only the estimator's spread, never its mean.

\section{Results: characterization}\label{sec:results}
All measurements use a single consumer GPU (NVIDIA RTX~5060~Ti, 8\,GB), qiskit~1.2.4,
qiskit-aer-gpu~0.15.1, cupy~14.1.1, with fixed seeds; raw data accompany the artifact.

\subsection{Closed-form variance reproduction}
We first confirm that the engine reproduces the closed forms \eqref{eq:varstd}--\eqref{eq:varproj}
in the canonical single-qubit dephasing scenario ($O=X$, $\ket{+}$), over $N=2\times10^4$
trajectories. Empirical means match $e^{-2\gamma t}$ throughout (e.g.\ $0.0456$ vs.\ $0.0498$ at
$\gamma t=1.5$). Empirical variances match the closed forms to within Monte-Carlo error: at
$\gamma t=1.5$, standard $0.9979$ vs.\ predicted $0.9975$ and projector $0.0485$ vs.\ predicted
$0.0473$---a $20.6\times$ reduction; at $\gamma t=0.1$, analog attains the lowest variance
($0.069$ vs.\ standard $0.329$, projector $0.148$), confirming its weak-noise advantage.

\subsection{Regime map}
\begin{figure}[!htbp]\centering
\includegraphics[width=\linewidth]{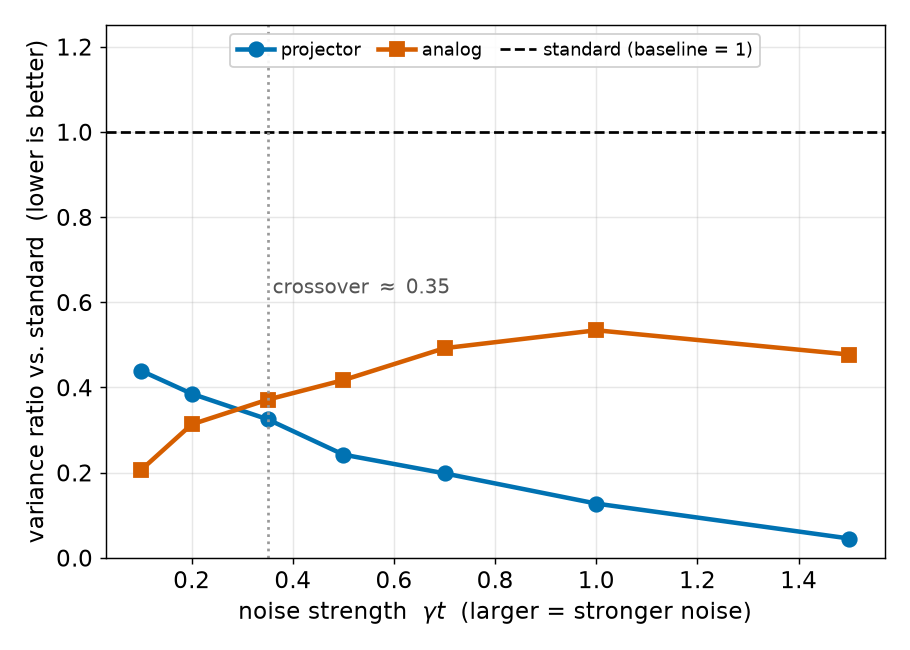}
\caption{Regime map of unraveling efficiency. The $x$-axis is the accumulated noise strength
$\gamma t$ (larger $=$ stronger noise); the $y$-axis is the per-trajectory variance of the
estimator \emph{relative to} the standard unraveling, so lower is better and the black dashed line
at $1$ is the standard baseline. Blue circles are projector, orange squares are analog (single
dephasing channel, $n=12$, $N=2\times10^3$). Both curves lie entirely below $1$, i.e.\ both
variance-reduced unravelings beat standard at every noise level. The projector ratio falls
monotonically to $0.045$ at $\gamma t=1.5$ (a $\sim\!22\times$ variance reduction), while analog is
lowest at weak noise; the two curves cross near $\gamma t\approx0.35$ (grey dotted). Practical
reading: use analog to the left of the crossover and projector to the right. Because deeper or
larger noisy circuits accumulate more $\gamma t$, they fall into the projector-favorable (right)
region.}
\label{fig:regime}
\end{figure}

Figure~\ref{fig:regime} and Table~\ref{tab:regime} map the per-trajectory variance ratio
(unraveling vs.\ standard) against noise strength $\gamma t$ at $n=12$, $N=2\times10^3$. Both
projector and analog beat standard everywhere. The projector ratio decreases monotonically from
$0.44$ at $\gamma t=0.1$ to $0.045$ at $\gamma t=1.5$ (i.e.\ $\sim\!5\%$ of the standard variance,
a $\sim\!22\times$ reduction), while the analog ratio rises from $0.21$ to $\sim\!0.48$. The two
cross near $\gamma t\approx0.35$ (analog $0.371$ vs.\ projector $0.325$): analog is preferable
below the crossover, projector above it.

\begin{table}[!htbp]
\caption{Regime map: variance ratio vs.\ standard ($n=12$, $N=2\times10^3$). Lower is better.}
\label{tab:regime}
\centering
\begin{tabular}{cccc}
\toprule
$\gamma t$ & $\mathrm{Var}_{\mathrm{proj}}/\mathrm{Var}_{\mathrm{std}}$ &
$\mathrm{Var}_{\mathrm{analog}}/\mathrm{Var}_{\mathrm{std}}$ & winner \\
\midrule
0.10 & 0.439 & 0.206 & analog \\
0.20 & 0.385 & 0.313 & analog \\
0.35 & 0.325 & 0.371 & $\approx$ crossover \\
0.50 & 0.242 & 0.417 & projector \\
0.70 & 0.198 & 0.492 & projector \\
1.00 & 0.127 & 0.534 & projector \\
1.50 & 0.045 & 0.477 & projector \\
\bottomrule
\end{tabular}
\end{table}

\subsection{Trajectory savings and scaling}
\begin{figure}[!htbp]\centering
\includegraphics[width=\linewidth]{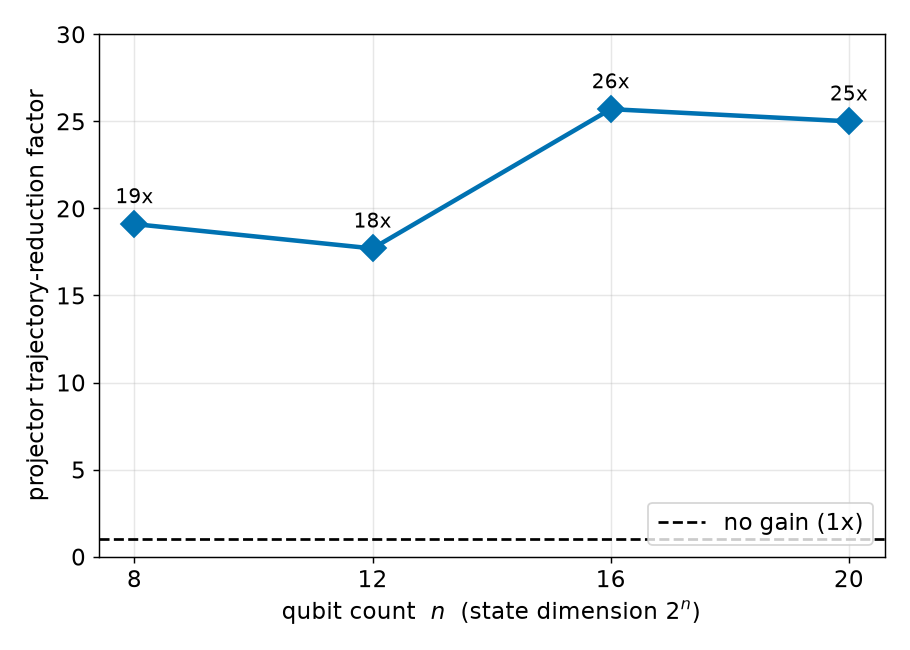}
\caption{Scaling of the trajectory-reduction factor. The $x$-axis is the qubit count $n$ (state
dimension $2^n$, i.e.\ $256$ to $\sim\!10^6$ amplitudes); the $y$-axis is how many times fewer
trajectories the projector unraveling needs than standard to reach the same target standard error
($\mathrm{SE}\le10^{-2}$, strong noise $\gamma t=1.5$). Each blue marker is a measured ratio
(Table~\ref{tab:scaling}); the black dashed line at $1$ marks ``no advantage''. The factor stays
in the $19$--$26\times$ band with no downward trend across $n=8$--$20$, so the benefit is a
statistical property of the unraveling rather than a small-system artifact. Statevector memory is
only $16.8$\,MB at $n=20$, four orders of magnitude below the $8$\,GB budget, indicating the method
is compute-bound here and that the same advantage is expected to persist to the $\sim\!30$-qubit
regime reachable on larger-memory GPUs.}
\label{fig:scaling}
\end{figure}

Table~\ref{tab:scaling} and Fig.~\ref{fig:scaling} report the number of trajectories to reach a
target standard error ($\mathrm{SE}\le10^{-2}$) for standard vs.\ projector, and the resulting
speedup, as $n$ grows in the strong-noise regime ($\gamma t=1.5$). Standard needs
$539$--$1186$ trajectories; projector needs only $22$--$62$, a speedup of $19.1\times$ ($n=8$) to
$25.7\times$ ($n=16$), essentially flat across the range. GPU memory for the statevector is
$16.8$\,MB at $n=20$---four orders of magnitude below the $8$\,GB ceiling---so the method is
compute-bound, not memory-bound, at these sizes; the advantage is a statistical property
independent of $n$, not a small-$n$ artifact.

\begin{table}[!htbp]
\caption{Scaling of trajectories-to-target ($\mathrm{SE}\le10^{-2}$, $\gamma t=1.5$, $N$ sampled).}
\label{tab:scaling}
\centering
\begin{tabular}{cccccc}
\toprule
$n$ & dim $2^n$ & $N_{\mathrm{std}}$ & $N_{\mathrm{proj}}$ & speedup & SV mem \\
\midrule
 8 & 256      & 1186 & 62 & 19.1$\times$ & 4\,KB \\
12 & 4096     &  785 & 44 & 17.7$\times$ & 64\,KB \\
16 & 65\,536  &  632 & 25 & 25.7$\times$ & 1\,MB \\
20 & 1\,048\,576 & 539 & 22 & 25.0$\times$ & 16.8\,MB \\
\bottomrule
\end{tabular}
\end{table}

\subsection{Entangled states: observable choice matters}
For a GHZ state the single-qubit observable $\expv{X_i}$ vanishes identically, so it carries no
signal. The physically meaningful quantity is the GHZ stabilizer $X^{\otimes n}$, which measures
the $\ket{0}^{\otimes n}$--$\ket{1}^{\otimes n}$ coherence, decays as $e^{-2n\gamma t}$, and
anticommutes with each dephasing generator $Z_i$. With this observable ($n=10$) the projector
estimator is again Bernoulli with $\mathrm{Var}_{\mathrm{proj}}=m(1-m)$ for mean $m$: measured
means $0.3645,0.1360,0.0380$ track the exact $e^{-2n\gamma t}=0.368,0.135,0.050$ at
$\gamma t=0.05,0.10,0.15$, and the projector speedup grows $3.6\times\!\to\!8.4\times\!\to\!27.2\times$
as the effective noise $n\gamma t$ increases. The earlier ``no benefit for GHZ'' is thus an
observable-choice artifact, not a limitation of the method.

\subsection{Coherent erosion: an applicability map}
\begin{figure}[!htbp]\centering
\includegraphics[width=\linewidth]{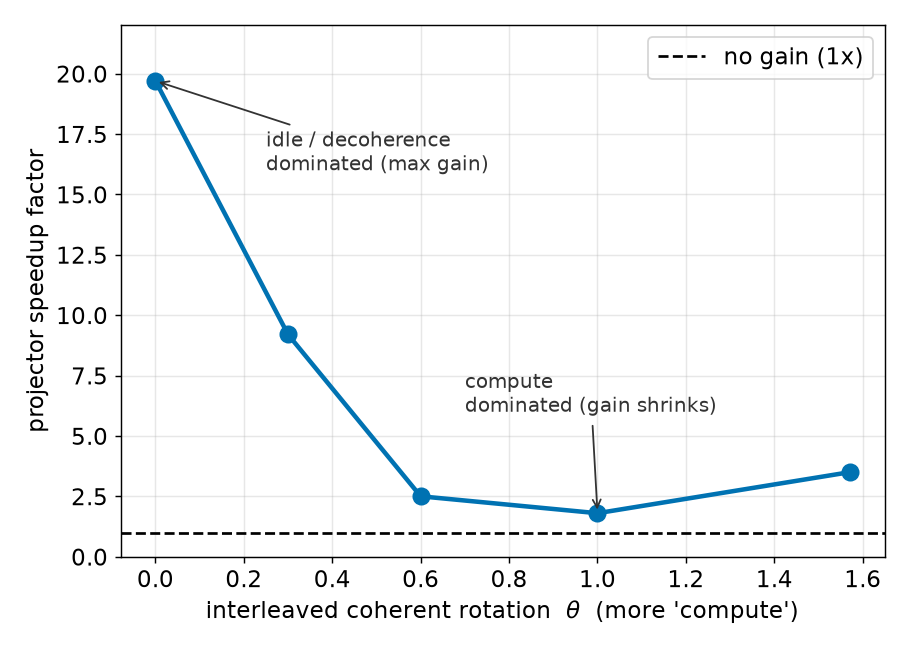}
\caption{Applicability map: coherent dynamics erodes the projector advantage. The $x$-axis is the
angle $\theta$ of $R_Y(\theta)$ rotations interleaved between noise windows---a proxy for how much
coherent ``compute'' the circuit performs relative to idle decoherence; the $y$-axis is the projector
trajectory-reduction factor ($n=8$, $\gamma t=1.5$). At $\theta=0$ (pure idle/dephasing, the
absorbing-window ideal) the speedup is largest, $19.7\times$; it falls monotonically to $1.8\times$
at $\theta=1.0$ as $R_Y$ (which does not commute with the measured $X$) repeatedly rotates collapsed
states back out of the absorbing subspace, with a partial rebound to $3.5\times$ at $\theta=\pi/2$.
Crucially the estimator mean remains unbiased at every $\theta$ (it always tracks the exact value);
only the \emph{size} of the variance advantage changes. The practical message is that projector
unraveling pays off most for decoherence/idle-dominated workloads (e.g.\ quantum-memory and
long-delay circuits), which is precisely where noisy-simulation cost is otherwise highest.}
\label{fig:erosion}
\end{figure}

The projector advantage relies on the absorbing window, which coherent evolution can reopen.
Interleaving $R_Y(\theta)$ rotations between noise windows ($n=8$, $\gamma t=1.5$; $R_Y$ does not
commute with $X$) erodes it monotonically: the projector speedup falls $19.7\times$ ($\theta=0$,
idle) $\to 9.2\times$ ($0.3$) $\to 2.5\times$ ($0.6$) $\to 1.8\times$ ($1.0$), with a partial
rebound to $3.5\times$ at $\theta=\pi/2$ (Fig.~\ref{fig:erosion}). The estimator mean stays
unbiased throughout ($-0.14\le\expv{X}\le0.05$ tracks the exact value). Projector unraveling is
therefore most powerful in decoherence/idle-dominated regimes---an applicability map for
practitioners, not a correctness limit.

\section{Head-to-head with Qiskit-Aer and the integration gap}\label{sec:aer}
\subsection{Benchmark}
\begin{figure}[!htbp]\centering
\includegraphics[width=\linewidth]{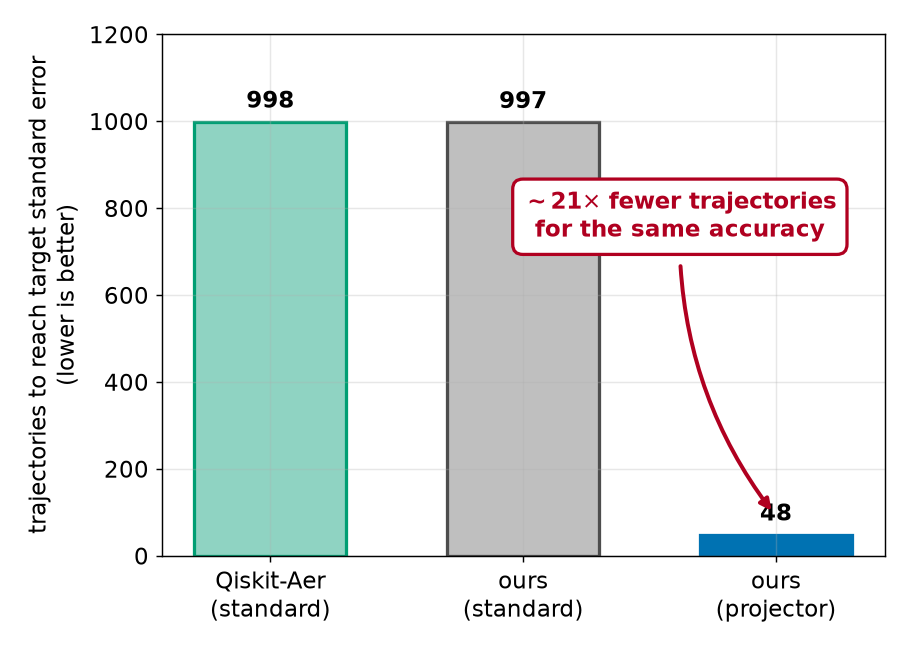}
\caption{Head-to-head against Qiskit-Aer. Bar height is the number of trajectories (equivalently
shots) needed to reach the target standard error $\mathrm{SE}\le10^{-2}$ on the observable
$\frac1n\sum_i X_i$; lower is better. Conditions: $n=10$, strong noise $\gamma t=1.5$, exact mean
$e^{-2\gamma t}=0.0498$, all three methods reproduce it (Table~\ref{tab:aer}). Left (green):
Qiskit-Aer's production \texttt{batched\_shots\_gpu} standard unraveling, $998$. Middle (black): our
engine's standard unraveling, $997$---statistically indistinguishable from Aer, which certifies that
our engine is a faithful, apples-to-apples baseline and that the improvement is due to the
unraveling, not the implementation. Right (blue): our projector unraveling, $48$---a $\sim\!21\times$
reduction (red arrow) at identical accuracy. Wall-clock is discussed separately
(Section~\ref{sec:aer}): the saving is in \emph{trajectory count}, an engine-independent statistical
quantity that a compiled implementation would convert into proportional wall-clock.}
\label{fig:aer}
\end{figure}

Table~\ref{tab:aer} and Fig.~\ref{fig:aer} compare against Aer's \texttt{batched\_shots\_gpu}
at $n=10$, $\gamma t=1.5$ (exact mean $e^{-2\gamma t}=0.0498$), target $\mathrm{SE}\le10^{-2}$.
All three methods recover the exact mean ($0.0497$ Aer, $0.0452$ ours-standard, $0.0499$
ours-projector). Our standard unraveling has the \emph{same} per-sample spread as Aer
($0.3158$ vs.\ $0.3158$) and needs $997$ vs.\ Aer's $998$ trajectories---statistically identical,
establishing an apples-to-apples baseline. Projector cuts the per-sample standard deviation to
$0.0695$ and needs only $48$ trajectories, a $20.8\times$ reduction versus Aer.

Wall-clock is reported honestly. Aer's optimized C++/CUDA engine is fastest today
($2.7\times10^4$ trajectories/s vs.\ our Python prototype's $5.7\times10^2$--$2.4\times10^2$/s),
so Aer reaches the target in $36.7$\,ms versus our projector's $199.8$\,ms. Within our own engine,
however, projector already beats standard on wall-clock ($199.8$ vs.\ $1747.6$\,ms, $8.7\times$),
purely from the $20.8\times$ fewer trajectories partially offset by a $2.4\times$ per-trajectory
collapse cost. The trajectory saving is engine-independent; ported into Aer's compiled path it
would carry directly to wall-clock, as we discuss next.

\begin{table}[!htbp]
\caption{Head-to-head at $n=10$, $\gamma t=1.5$ (exact mean $0.0498$; target $\mathrm{SE}\le10^{-2}$).}
\label{tab:aer}
\centering

\begin{tabular}{@{}lcccr@{}}
\toprule
method & mean & std & $N_{\mathrm{tgt}}$ & time \\
\midrule
Aer (standard)   & 0.0497 & 0.3158 & 998 & 36.7\,ms \\
ours (standard)  & 0.0452 & 0.3158 & 997 & 1747.6\,ms \\
ours (projector) & 0.0499 & 0.0695 & \textbf{48} & 199.8\,ms \\
\bottomrule
\end{tabular}
\end{table}

\subsection{Why Aer cannot express this today}
We attempted to inject projector unraveling through Aer's public API by supplying the dephasing
channel in the projector Kraus decomposition
$\{\sqrt{1-2p}\,\mathbb{1},\sqrt{2p}\,\Pi_+,\sqrt{2p}\,\Pi_-\}$ (identical CPTP map to the
standard $\{\sqrt{1-p}\,\mathbb{1},\sqrt{p}\,Z\}$). Aer produced the correct mean but the
\emph{standard} variance---no reduction. Inspection shows Aer stores the error as a single
\texttt{kraus} instruction and, at apply time, reconstructs a \emph{canonical} Kraus set from
the channel (the Choi eigen-Kraus, i.e.\ the orthogonal Pauli set for dephasing), discarding the
user-supplied non-canonical decomposition. Thus variance-reduced unravelings cannot be delivered
through the API: Aer preserves the channel, not the unraveling.

\subsection{A minimal integration path}
Critically, the required machinery already exists: for a genuinely non-unital channel
(amplitude damping) Aer's statevector \texttt{kraus} sampler performs proper Born-rule collapse
(per-trajectory $\expv{Z}$ is bimodal, $29.6\%$ collapsed). The minimal change is therefore to
(i) let a \texttt{QuantumError} carry a ``preserve-unraveling'' flag so the supplied Kraus set
survives to apply time, (ii) sample $K_i$ by $\|K_i\psi\|^2$ over that set rather than the
canonical one, and (iii) expose a per-trajectory expectation accumulator. This is a localized
addition that would make variance-reduced unravelings available in production and realize the
$21\times$ saving on Aer's optimized engine.

\section{Related work}
\subsection{Statevector and density-matrix simulators}
High-performance statevector simulators scale ideal circuits across CPUs, GPUs, and clusters:
qHiPSTER/Intel-QS \cite{intelqs2020}, QuEST \cite{quest2019}, the $45$-qubit supercomputer
simulation of H\"aner and Steiger \cite{haner2017}, NVIDIA cuQuantum/cuStateVec
\cite{cuquantum2023}, and Qiskit-Aer \cite{qiskit_aer} with cache blocking
\cite{cacheblocking2020}. These optimize \emph{throughput} for ideal (and, for the
density-matrix method, small-$n$ noisy) simulation. Our work is orthogonal: we reduce the
\emph{statistical} cost of the trajectory method rather than the per-amplitude cost, and we do
not compete on raw kernel speed.

\subsection{GPU circuit partitioning and multi-GPU scaling}
Atlas \cite{atlas_sc24}, quEStab \cite{questab_ics25}, and multi-GPU network-performance studies
\cite{multigpu_network2025} partition a single large statevector (or stabilizer state) across GPUs
and nodes, and analyze communication overhead. This axis---scaling \emph{ideal} circuits to more
qubits---is complementary to ours (fewer \emph{trajectories} for \emph{noisy} circuits) and is the
natural setting for a future multi-GPU extension of our method. GPU multi-shot batching
\cite{efficient_multishot2023} and ROCm support \cite{rocm_qiskit2024} are complementary engine
features that our technique would ride on once integrated.

\subsection{Trajectory unravelings and open-system methods}
The MCWF/quantum-trajectory method has a long history in quantum optics
\cite{dalibard1992,molmer1993,plenio1998,breuer_petruccione,daley2014}; the freedom to choose an
unraveling is classical, but exploiting it for variance reduction in circuit simulation is recent.
The tensor jump method \cite{tjm2026} introduces the projector and analog unravelings we adopt,
with sparse Pauli-Lindblad noise \cite{sparse_pauli_lindblad2023} on MPS backends
\cite{vidal2003,cirac2021mps}. We port these unravelings to a GPU dense statevector---a different
representation with different cost trade-offs (exact amplitudes, no bond-dimension truncation, but
$2^n$ memory)---and, uniquely, analyze their integration into a production simulator.

\subsection{Approximate noisy simulation and error mitigation}
A complementary line approximates noisy dynamics with tensor networks: locally purified density
operators \cite{lpdo2023}, matrix-product density operators \cite{mpdo_prr,mpdo_tomography2025},
and noise-induced entanglement bounds enabling efficient classical simulation
\cite{noh2020,zhou2020}. These trade exactness for compression and, like \cite{tjm2026}, are
CPU/tensor-network methods absent from production GPU tooling. Finally, error-mitigation
techniques such as zero-noise extrapolation and probabilistic error cancellation
\cite{temme2017,sparse_pauli_lindblad2023} consume large numbers of noisy expectation-value
estimates, which is precisely the workload whose per-estimate trajectory cost our method reduces.

\section{Limitations}\label{sec:limitations}
Three limitations bound the present claims. \emph{(i) Prototype wall-clock.} Our engine is a Python
(cupy) prototype; its absolute wall-clock trails Aer's compiled C++/CUDA engine, so the $21\times$
saving is demonstrated in \emph{trajectory count} and becomes wall-clock only after a CUDA-kernel or
Aer-C++ implementation. \emph{(ii) Scale.} Measurements use a single $8$\,GB consumer GPU with $n\le20$
swept; larger $n$ is projected, not yet measured. \emph{(iii) Regime dependence.} The projector
advantage is largest for decoherence/idle-dominated circuits and shrinks when strong coherent dynamics
is interleaved (Fig.~\ref{fig:erosion}); the estimator remains unbiased throughout, but the speedup is
not universal. We also restrict attention to Pauli-Lindblad noise, for which jump rates are
state-independent; non-Pauli channels require the general norm-dependent sampling of \eqref{eq:jumpprob}.

\section{Future applications and research directions}\label{sec:future}

\subsection{Scaling to datacenter GPUs}
Because the trajectory reduction is a statistical property, it is hardware-independent; a larger GPU buys
\emph{reach} rather than a larger factor. An $80$\,GB A100/H100 admits $n\!\approx\!32$ statevector
trajectories---a regime where the exact density matrix ($4^n$) is impossible ($n\!\approx\!16$)---so the
same $\sim\!21\times$ advantage would apply to $30$-qubit noisy circuits. Combined with a compiled kernel
(memory-bandwidth gain of $\sim\!5$--$10\times$ over the consumer GPU used here), the compounded
improvement over a standard-unraveling baseline is expected to be one to two orders of magnitude.

\subsection{Target applications}
The method is most valuable wherever many \emph{noisy expectation values} are needed:
(a) variational algorithms (VQE/QAOA) under realistic device noise \cite{cerezo2021vqa,bharti2022nisq};
(b) error-mitigation pipelines---zero-noise extrapolation and probabilistic error cancellation
\cite{temme2017,kandala2019mitigation,endo2021mitigation,sparse_pauli_lindblad2023}---whose cost is
dominated by repeated noisy estimates; (c) device characterization and noise-model validation, i.e.\
fitting Lindblad generators to measured data \cite{georgopoulos2021modeling,nation2023suppressing};
(d) quantum-memory and idle-dominated protocols, which sit exactly in the projector-favorable regime
(Fig.~\ref{fig:erosion}); and (e) classical-simulation benchmarks used to assess quantum-advantage
claims \cite{arute2019supremacy,pan2022sycamore,kim2023utility}.

\subsection{Algorithmic extensions}
Natural next steps include: \emph{adaptive unraveling selection}, in which the simulator chooses
projector, analog, or standard per channel and per circuit window using the measured crossover
(Fig.~\ref{fig:regime}) as a runtime policy; extension beyond Pauli-Lindblad to amplitude damping and
general non-unital channels, where Aer's Born-rule collapse already applies; \emph{hybrid schemes} that
combine variance-reduced unravelings with MPS/MPDO representations
\cite{tjm2026,verstraete2004mpdo,mpdo_prr,lpdo2023} so that both the statistical and the representational
cost are reduced simultaneously; and importance-sampling or control-variate estimators layered on top of
the unraveling choice.

\subsection{Systems directions}
Beyond the minimal Aer change of Section~\ref{sec:aer}, we see three systems avenues: multi-GPU partitioning
of each trajectory for $n>33$ \cite{atlas_sc24,questab_ics25,multigpu_network2025}; cross-vendor
evaluation on AMD ROCm \cite{rocm_qiskit2024} to test whether the advantage is portable across
architectures; and integration with cuStateVec \cite{cuquantum2023}, noting that Aer currently disables
cuStateVec on the batched-shots path---a seam where a variance-reduced, collapse-aware kernel could be
particularly valuable.

\section{Note on AI-assisted development methodology}\label{sec:ai}
The engine, benchmark harness, and reproduction scripts in this work were developed with substantial
assistance from an AI coding agent (Claude Code), following a methodology the author has previously
applied and reported in two scientific-software optimization studies: a CUDA-graph-based optimization of
the micromagnetic simulator MuMax3 \cite{you2026mumax}, and a systematic large-language-model-assisted
optimization of the LAMMPS molecular-dynamics simulator \cite{you2026lammps}. In both cases the
productive pattern was the same one used here: the AI agent is directed to (i) \emph{measure first},
establishing a correctness oracle and a reproducible benchmark before any optimization; (ii) attack a
\emph{specific, hypothesis-driven} bottleneck rather than performing undirected refactoring; and
(iii) report negative results faithfully---in this work, the finding that the projector unraveling
\emph{cannot} be injected through Qiskit-Aer's public API (Section~\ref{sec:aer}) emerged from such an
attempt and became a central contribution. We record this both for transparency and because the
cross-domain reuse of a GPU-optimization methodology (micromagnetics $\rightarrow$ molecular dynamics
$\rightarrow$ quantum-circuit simulation) is itself a transferable result. All quantitative claims
reported here were verified by executing the accompanying scripts; no performance number is
AI-generated.

\section{Conclusion}
Variance-reduced trajectory unravelings, previously demonstrated only on CPU tensor networks,
provide a $\sim\!21\times$ reduction in trajectory count for GPU dense-statevector noisy
simulation, with a clean noise-strength regime map and unbiased estimates validated against the
exact density matrix. We further show that production Qiskit-Aer cannot express these unravelings
because it canonicalizes noise channels at apply time, and we specify the minimal change that
would unlock the technique. The result is a statistically grounded, production-relevant path to
cheaper noisy quantum-circuit simulation.

\section*{Acknowledgment}
This work was supported by the National Research Foundation of Korea (NRF)
(No. RS-2025-25463492) and the Strategic Research Program under the DGIST R\&D Program
(26-SR-01) of the Ministry of Science, ICT, and Future Planning.

\emph{Disclosure of AI usage.} The author discloses that an AI
coding assistant, Claude Code (Anthropic), was used under the author's direction to implement the
simulation and analysis scripts, execute the benchmark runs, and draft expository text. It did not
originate the research question, hypotheses, interpretation, or conclusions, and is not an author.
All reported numbers were produced by executing the accompanying scripts and were verified by the
author, who takes full responsibility for the manuscript. The underlying methodology is described in
Section~\ref{sec:ai} and in \cite{you2026mumax,you2026lammps}.

\emph{Data and code availability.} Reproduction scripts, raw data, environment manifests, and
figure-generation code accompany this manuscript as supplementary material and are available
at \url{https://github.com/mirryou-maker/qiskit-variance-reduced-unraveling}.

\bibliographystyle{unsrt}
\bibliography{references}

\end{document}